\title{Level sets of the lapse function in static GR}

\author{Carla Cederbaum\footnote{Email: cederbaum@math.uni-tuebingen.de} \\
	Mathematisches Institut\\
        Universit\"at T\"ubingen\\
        Auf der Morgenstelle 10\\
        72076 T\"ubingen, Germany}

\date{\today}

\documentclass[11pt,english]{article}
\usepackage[]{graphicx}

\begin{document}
\maketitle

\begin{abstract}
We present a novel physical interpretation of the level sets of the (canonical) lapse function in static isolated general relativistic spacetimes. Our interpretation uses a notion of \emph{constrained test particles}. It leads to a definition of gravitational force on test particles and to a previously unknown uniqueness result for the lapse function.

In Section \ref{photo}, we discuss \emph{photon spheres} in static isolated relativistic spacetimes and relate them to the level sets of the lapse function.

\end{abstract}

\section{Static isolated relativistic spacetimes}\label{intro}
Static isolated relativistic spacetimes have been studied from a number of perspectives including regularity, asymptotics, construction of explicit solutions etc. They serve as models of static isolated star and black hole configurations. The standard example is the Schwarzschild family of spacetimes
\begin{eqnarray}
ds^2&=&-N^{2}c^{2}dt^{2}+\frac{1}{N^{2}}dr^{2}+r^{2}d\Omega^{2}, \label{schwarz}\\
N&=&\left(1-\frac{2M}{r}\right)^{\frac{1}{2}},\label{schwarzlapse}
\end{eqnarray}
modeling the vacuum exterior region of a black hole or spherically symmetric matter distribution. Here, $M=mGc^{-2}$ is the mass parameter and $d\Omega^{2}$ denotes the standard metric of the round sphere.

$N$ is called the \emph{(canonical) lapse function}. For $M=m=0$, we find the Minkowski spacetime as a special case in the Schwarzschild family.

More generally, a static relativistic spacetime is given by a smooth Lorentzian metric $ds^{2}$ which has a smooth timelike Killing vector field $X=\partial_{t}$ which is hypersurface-orthogonal with respect to the induced Levi-Civit\'a connection $\nabla$:
\begin{equation}\label{hyporth}
X_{\left[\alpha\right.}{\nabla}_{\beta} X_{\left.\gamma\right]}=0.
\end{equation}

\begin{figure}[h]
\begin{center}
\includegraphics[scale=0.3]{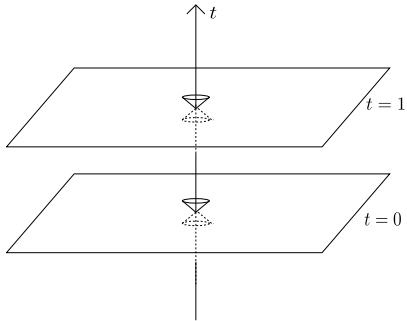}
\caption{The time-slices of a canonically decomposed static spacetime.}
\label{timeslice}
\end{center}
\end{figure}

It follows that the \emph{(canonical) lapse function} $N:=\sqrt{-ds^{2}(X,X)}$ is positive and time-independent. The metric locally splits into a warped product
\begin{equation}\label{metric}
ds^{2}=-N^{2}c^{2}dt^{2}+g,
\end{equation}
where $g$ is a time-independent Riemannian metric on a time-slice $\lbrace{t=\mbox{const}}\rbrace$. All time-slices are isometric and extrinsically flat, cf.\ Figure \ref{timeslice}.

A static spacetime with metric $ds^{2}$ of the form (\ref{metric}) is said to model an \emph{isolated system} if
\begin{eqnarray}\label{asym}
g_{ij}&\to&\delta_{ij}\\\label{asymN}
N&\to&1
\end{eqnarray}
as $r\to \infty$ and if the topology of the time-slices is that of Euclidean $3$-space outside a ball. Thinking of the time-slices as only modeling the exterior region outside the (assumedly finitely extended) matter distribution and outside all black holes, let us additionally assume that $ds^{2}$ satisfies the vacuum Einstein equations. Making use of the time-translation symmetry, the vacuum Einstein equations reduce to the \emph{static vacuum equations}
\begin{eqnarray}\label{SME1}
N\mbox{Ric}&=&\nabla^{2}N\\\label{SME2}
\triangle N&=&0
\end{eqnarray}
on any and every time-slice. Here, $\mbox{Ric}$ denotes the Ricci curvature tensor, $\nabla^{2}$ the covariant Hessian, and $\triangle$ the covariant Laplacian corresponding to the $3$-dimensional Riemannian metric $g$ on the time-slices.

While the Riemannian metric $g$ captures the geometry of the time-slices, the lapse function $N$ incorporates the information of how we have to arrange or stack the (isometric) time-slices in order to reproduce the static spacetime, cf.\ Equation (\ref{metric}) and Figure \ref{timeslice}. 

\section{Physical interpretation of the lapse function}\label{interpretation}
What is the physical significance of the lapse function $N$? Does it play a role similar to that of the Newtonian potential in static Newtonian gravity? Surprisingly, it (almost) does, and in a number of ways:
\begin{itemize}
\item {\bf Newtonian limit:} The \emph{pseudo-Newtonian potential} $U:=c^{2}\ln N$ converges to the Newtonian potential in the Newtonian limit (along any suitable family of static isolated spacetimes possessing a Newtonian limit in the sense of Ehlers' frame theory), cf.\ \cite{Ehlers:1989} and \cite{Cederbaum:2011}.
\item {\bf Asymptotic behavior:} The pseudo-Newtonian potential $U=c^{2}\ln N$ asymptotically behaves like
\begin{equation}\label{decay}
U=-\frac{mG}{r}+\mathcal{O}(\frac{1}{r^{2}})
\end{equation}
as $r\to\infty$, where $m$ is the ADM-mass of the system, cf.\ \cite{Kennefick:1995}. Moreover, the next order term in the asymptotic expansion of $U$ corresponds to a center of mass term of the same form as in the Newtonian setting
\begin{equation}
U=-\frac{mG}{r}-\frac{mG\vec{z}\cdot\vec{x}}{r^{3}}+\mathcal{O}(\frac{1}{r^{3}}),
\end{equation}
cf.\ \cite{Cederbaum:2011}. Both asymptotic decay statements are made precise using weighted Sobolev spaces.
\item {\bf Equipotential surfaces:} Relativistic test particles constrained to a given $2$-surface $\Sigma^{2}$ generically accelerate along $\Sigma^{2}$ unless $\Sigma^{2}$ is a level set of the lapse function $N$ and thus also of the pseudo-Newtonian potential $U=c^{2}\ln N$. It is thus justified to call those level sets \emph{equipotential}. We will describe this phenomenon in more detail in Section \ref{test}.\\[-1ex]
\item {\bf Uniqueness of lapse:} $N$ and thus also $U=c^{2}\ln N$ are uniquely determined outside the matter distribution by the spatial metric $g$ -- and everywhere in a given time-slice by the metric $g$ and the matter variables induced on any time-slice by the energy momentum tensor. We will sketch how this follows from the equipotential nature of the level sets in Section \ref{unique}. More details as well as two alternative proofs can be found in \cite{Cederbaum:2011} and \cite{Cederbaum:2013b}.
\item {\bf Degrees of freedom:} Introducing asymptotically flat wave-harmonic coordinates $(x^{i})$ on a time-slice, $N$ (or, alternatively, $U=c^{2}\ln N$) together with the coordinates uniquely determine the components $g_{ij}$ of the spatial metric $g$ of a given static isolated spacetime. So, in this sense, static isolated spacetimes have four degrees of freedom, namely those corresponding to the choice of $(N,x^{1},x^{2},x^{3})$. This coincides with static isolated Newtonian gravity, where the degrees of freedom are given by the Newtonian potential and three Galilei coordinates.
\end{itemize}

\section{Equipotential surfaces and gravitational force}\label{test}
\begin{figure}[h]
\begin{center}
\includegraphics[scale=0.3]{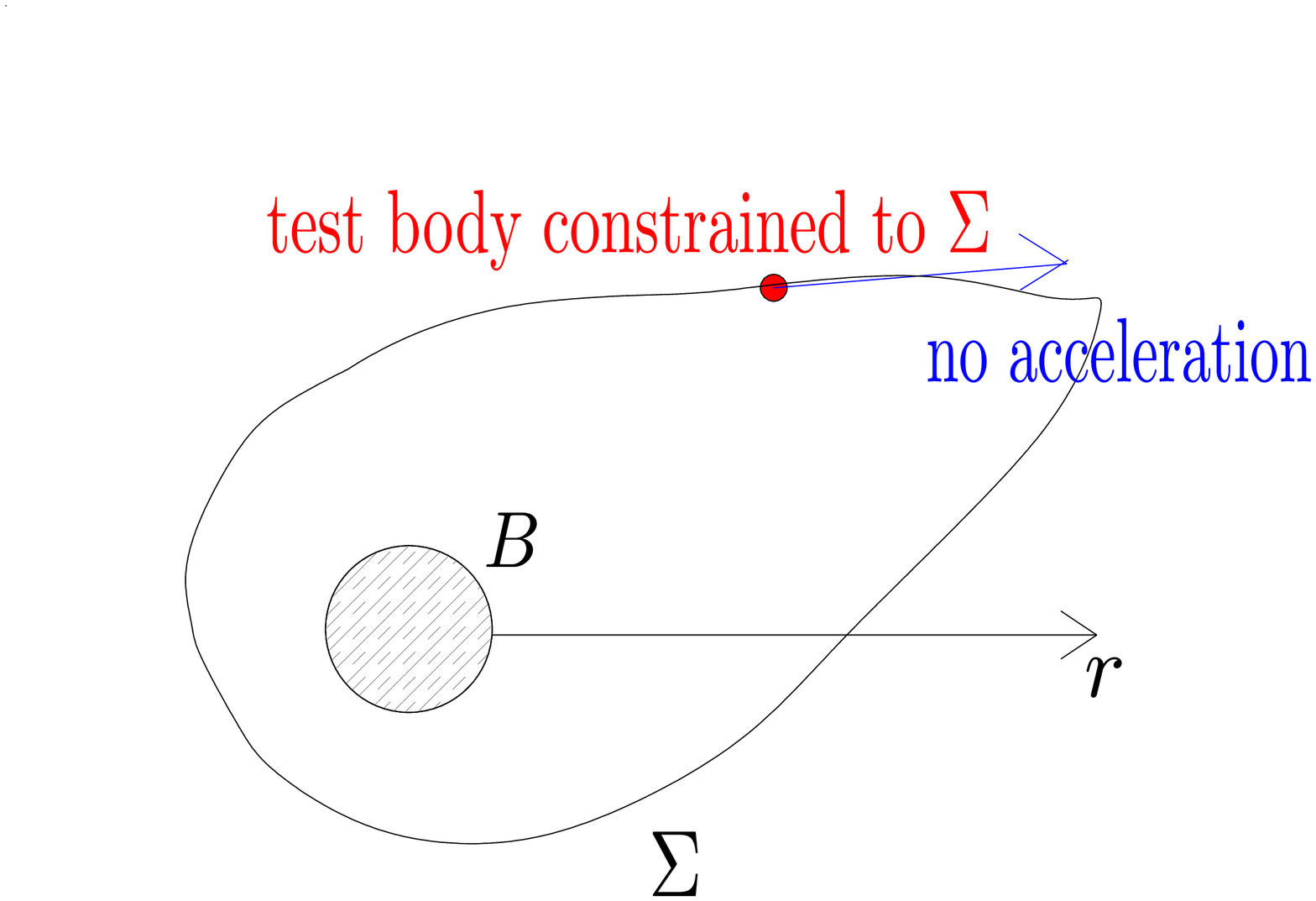}
\caption{A (Newtonian) test body constrained to a surface $\Sigma^{2}$.}
\label{levelset}
\end{center}
\end{figure}

In static Newtonian gravity, the (negative) gradient of the Newtonian potential defines the force $\vec{F}$ on a unit mass test body. This has a well-known consequence for the \emph{equipotential} or level set surfaces: if a test body is constrained into one of the level set surfaces, the gravitational force $\vec{F}$ is perpendicular to the path of the test body and does thus not have a tangential component. Hence the test body does not tangentially accelerate along any level set surface, see Figure \ref{levelset}. If, on the other hand, a test body is constrained to an arbitrary surface $\Sigma^{2}$, it will in general accelerate due to the tangential component of the force along $\Sigma^{2}$.

Surprisingly, the ``same'' is true for the level sets of the lapse function $N$ in a static isolated spacetime. To see that, we need to replace the constrained Newtonian test bodies in the above picture by ``constrained (relativistic) test particles''. In order to define these, recall that a timelike curve $\mu(\tau)$ is called a \emph{(freely falling relativistic) test particle} if it is a critical point of the time functional $T\left[\mu\right]$ given by
\begin{equation}\label{T}
T\left[\mu\right]:=\int_{\tau_{0}}^{\tau_{1}} \vert\dot{\mu}(\tau)\vert\,d\tau.
\end{equation}
In the context of static isolated spacetimes, the warped product structure of the metric captured by (\ref{metric}) allows us to think of a timelike curve $\mu(\tau)$ as having temporal and spatial components such that
\begin{equation}
\mu(\tau)=(t(\tau),x(\tau)).
\end{equation}
Accordingly, let us say that a timelike curve $\mu(\tau)$ is \emph{constrained to a surface $\Sigma^{2}\subset\lbrace t=\mbox{const}\rbrace$} if $x(\tau)\in\Sigma^{2}$ for all $\tau$. In other words, we ask the curve $\mu(\tau)$ to stay within the warped $3$-dimensional timelike cylinder $(-\infty,\infty)\times\Sigma^{2}$ ``over'' $\Sigma^{2}$ in the spacetime. We then define a \emph{constrained (relativistic) test particle} to be a timelike curve $\mu(\tau)$ which is a critical point of the time functional $T\left[\mu\right]$ as in (\ref{T}) but subject to the constraint that it (and all its competitors in the variation of $T\left[\mu\right]$) is constrained to a given surface $\Sigma^2\subset\lbrace t=\mbox{const}\rbrace$. 

Equipped with this notion of constrained test particles, let us say that a closed surface $\Sigma^{2}$ sitting in a time-slice $\lbrace t=\mbox{const}\rbrace$ of a static isolated spacetime is an \emph{equipotential surface}  if {\bf every} test particle constrained to $\Sigma^{2}$ is a geodesic in $\Sigma^{2}$ with respect to the induced $2$-metric, i.e.\ does not accelerate within $\Sigma^{2}$, see again Figure \ref{levelset}. This is a geometrized analog of the Newtonian notion of equipotential surfaces.

A straightforward analysis of the Euler-Lagrange equations of the functional $T\left[\mu\right]$ subject to the constraint that $x(\tau)\in\Sigma^{2}$ for all $\tau$ shows that $\Sigma^2$ is an equipotential surface in a given static isolated spacetime if and only if $\Sigma^{2}$ is a level set of the lapse function $N$. Thus, in their effect on constrained test particles, the level sets of the lapse function $N$ in static isolated spacetimes play precisely the same role as those of the Newtonian potential in static isolated Newtonian gravity.

\subsection{Gravitational force}
In Newtonian gravity, the equipotential property of the level sets of the potential stems from the fact that the (negative) gradient of the potential determines the gravitational force per unit mass. It is thus tempting to define gravitational force per unit mass as the (negative) gradient of the lapse function $N$ in our context. Our Newtonian limit analysis\footnote{cf.\ Section \ref{interpretation} for a sketch of and \cite{Cederbaum:2011} for more details on the Newtonian limit analysis of static isolated spacetimes.} however suggests that we should replace the lapse function $N$ by the pseudo-Newtonian potential $U=c^{2}\ln N$. Observe that $U$ has the same level sets as $N$.

Moreover, it is more adequate\footnote{This follows from the Newtonian limit analysis combined with a more detailed study of the geometry of \emph{pseudo-Newtonian gravity}, cf.\ \cite{Cederbaum:2011}.} to take the gradient of $U$ not with respect to the induced spatial metric $g$ but instead with respect to the conformally equivalent Riemannian metric $\gamma:=N^{2}\,g$. We suggest to call $\gamma$ the \emph{pseudo-Newtonian metric} of the given static isolated spacetime.

Finally, it is important to realize that a notion of gravitational force constructed by the above equipotential principle will only apply to test particles\footnote{Our definition of gravitational force can be extended to general timelike curves that are not necessarily geodesics. These can be interpreted as test particles that are subject to not only gravitational but also to non-gravitational forces (e.g.\ electro-magnetic ones).}. It will not readily\footnote{The author has some first ideas how to generalize the notion of gravitational force presented here to extended bodies. A second Newtonian law of motion for extended bodies, however, seems more difficult at the moment.} apply to extended bodies. This distinction is irrelevant in static Newtonian gravity due to the superposition principle for the linear Poisson equation satisfied by the Newtonian potential.

So let $\mu(\tau)=(t(\tau),x(\tau))$ be a test particle in a static isolated spacetime with metric $ds^{2}$ as in (\ref{metric}). Assume that $\mu$ has mass $m$ or that
\begin{equation}
ds^{2}(\frac{d\mu}{d\tau},\frac{d\mu}{d\tau})=-m^{2}
\end{equation}
for all $\tau$. We then define the \emph{gravitational force on the test particle $\mu(\tau)$} as 
\begin{equation}\label{force}
\vec{F}:=-m{\,^{\gamma}}\vec{\nabla}U,
\end{equation}
where ${\,^{\gamma}}\vec{\nabla}$ denotes the ($3$-dimensional) $\gamma$-covariant gradient. Furthermore, we define its ($3$-dimensional) \emph{acceleration}\footnote{The acceleration of the test particle is induced from the Lorentzian metric $ds^2$ and thus naturally refers to the spatial metric $g$ and not to the conformally transformed metric $\gamma$ because the definition of test particles relies on the ``dynamics'' of the spacetime. On the other hand, the definition of gravitational force uses the analogy of pseudo-Newtonian and Newtonian effects and thus should be formulated in pseudo-Newtonian terms.} by
\begin{equation}\label{acceleration}
\vec{a}:={\,^{g}}\vec{\nabla}_{\frac{dx}{d\tau}}\frac{dx}{d\tau}.
\end{equation}
Here, ${\,^{g}}\vec{\nabla}$ denotes the ($3$-dimensional) $g$-covariant gradient. A straightforward computation shows that with these definitions of gravitational force and acceleration, the second \emph{pseudo-}Newtonian law of motion
\begin{equation}
\vec{F}=m\vec{a}
\end{equation}
holds for all freely falling test particles of mass $m$, see \cite{Cederbaum:2011}.

\section{Uniqueness results}\label{unique}
In a static isolated spacetime, the Einstein constraint equations on every time-slice reduces to 
\begin{equation}\label{constraint}
R=0.
\end{equation}
We observe that the lapse function $N$ does not appear in this constraint equation. Using Choquet-Bruhat's local uniqueness theorem (see e.g.\\cite{Choquet:1980}) for the Einstein equations, this implies that the spacetime corresponding to a given spatial metric $g$ and lapse function $N$ is in fact \emph{independent}\footnote{However, we do assume that the lapse function exists in the first place.} of the lapse function $N$. It follows that $N$ is indeed {\bf unique}: Suppose there was a second lapse function $\widetilde{N}$ such that $\widetilde{N}\to1$ as $r\to\infty$. Then by the above, the levels of $N$ and $\widetilde{N}$ would be detected by constrained test particles. Those only depend on the Lorentzian metric $ds^{2}$ which we have seen to be independent of the lapse function whatsoever. Thus the level sets of $N$ and $\widetilde{N}$ coincide so that
$\widetilde{N}=f\circ N$ for some real valued function $f$. The static vacuum equations (\ref{SME1}), (\ref{SME2}) applied to $N$ and $\widetilde{N}$ then imply
\begin{eqnarray}\nonumber
0&=&\triangle \widetilde{N}=\triangle(f\circ N)\\\nonumber
&=&f''\circ N\,\Vert{\nabla} N\Vert_{g}^2+f'\circ N \,{\triangle N}\\\nonumber
&=&f''\circ N\,\Vert{\nabla} N\Vert_{g}^2
\end{eqnarray}
so that $f''= 0$ and thus $\widetilde{N}=\alpha N + \beta$ for some real numbers $\alpha,\beta$. Moreover,
\begin{eqnarray}\nonumber
\nabla^2\widetilde{N}&=&\widetilde{N}\,\mbox{Ric}\\\nonumber
&=&(\alpha N +\beta)\,\mbox{Ric}\\\nonumber
&=&\alpha\,{\nabla}^2 N+\beta\,\mbox{Ric}\\\nonumber
&=&{\nabla}^2 \widetilde{N}+\beta\,\mbox{Ric}.
\end{eqnarray}
If $g$ is not everywhere flat, this implies $\beta=0$. Finally $N,\widetilde{N}\to 1$ as $r\to\infty$ leads to $\alpha=1$ so that $N= \widetilde{N}$.

We interpret this result as saying that ``there is only one way of synchronizing time at different locations in a geometrostatic spacetime such that one sees staticity'' just as, for a Riemannian geodesic, ``there is only one way of walking along a geodesic such that one does not accelerate (up to affine re-parametrizations)''. The affine freedom of the parameter along the geodesic does not make an appearance in the static isolated spacetime picture because we fixed the lapse function to asymptotically converge to $1$ at spacelike infinity and therewith fixed the time unit.  

\section{Photon spheres}\label{photo}
It is well-known that the Scharzschild spacetime (\ref{schwarz}), (\ref{schwarzlapse}) with positive mass parameter $M=mGc^{-2}$ possesses a so-called \emph{photon sphere} at $r=3M$. This is to say that photons (aka null geodesics) initially tangent to the timelike cylinder $(-\infty,\infty)\times\lbrace r=3M\rbrace$ remain tangent to it or that ``photons get caught in the sphere $\lbrace r=3M\rbrace$''. Moreover, each photon's energy and frequency is constant in time (as observed by the static observers $N^{-1}\partial_{t}$). This is a very interesting phenomenon. It has proved relevant for understanding questions related to dynamical stability and to gravitational lensing\footnote{See \cite{Cederbaum:2013b} for more information.}.

As we have seen above, the positive mass Schwarzschild spacetimes are prime examples of static isolated spacetimes. It is thus natural to ask whether more general static isolated spacetimes can also possess photon spheres or whether the phenomenon is restricted to the spherically symmetric Schwarzschild case. To the best knowledge of the author, this question was first raised in \cite{CVE:2001}.

To study this question, let us define\footnote{See \cite{CVE:2001}, \cite{Perlick:2005}, and \cite{Cederbaum:2014} for the origins of this definition.} a \emph{photon surface} in a static isolated spacetime with metric $ds^{2}$ of the form (\ref{metric}) as a surface $\Sigma^{2}\subset\lbrace t=\mbox{const}\rbrace$ such that {\bf every} photon (i.e.\ null geodesic) initially tangent to the cylinder $(-\infty,\infty)\times\Sigma^{2}$ remains tangent to it. A photon surface possessing the property that the tangential photons have constant energy and frequency along the photon surface (in the eyes of the static observers) will be called a \emph{photon sphere}. It turns out that this property is equivalent to constancy of the lapse function $N$ along the photon surface, see Lemma 2.7 in \cite{Cederbaum:2014}.

An analysis\footnote{cf.\ \cite{Cederbaum:2014} for an exposition of this analysis.} of the null geodesic equation combined with the Gau{\ss}-Codazzi-Mainardi equations and the static vacuum equations (\ref{SME1}), (\ref{SME2}) shows that a photon sphere $\Sigma^{2}$ in a static isolated spacetime must necessarily have constant mean curvature (or \emph{expansion}). Furthermore, a photon sphere $\Sigma^{2}$ has constant (intrinsic) Gau{\ss} curvature and the normal derivative of the lapse function $\nu(N)$ must also be constant along $\Sigma^{2}$.

In \cite{Cederbaum:2014}, the author shows that a static vacuum isolated spacetime with metric as in (\ref{metric}) possessing a -- connected -- photon sphere $\Sigma^{2}$ must be isometric to a Schwarzschild spacetime outside the photon sphere under a mild technical condition. This gives a partial answer to the question raised above; for a priori disconnected photon spheres, the issue will be addressed in \cite{cedergal}, together with other results on photon surfaces. 

This theorem can be interpreted as a photon sphere uniqueness theorem; its proof in fact mimics Israel's proof of static black hole uniqueness  \cite{Israel:1967}. It allows to identify a given static vacuum isolated spacetime as in fact being a Schwarzschild spacetime whenever it is known to possess a photon sphere.

More precisely, if $\Sigma^{2}\subset\lbrace t=\mbox{const}\rbrace$ is a photon sphere then we can\footnote{making the same mild technical assumption as Israel \cite{Israel:1967} that the lapse function foliates the region exterior to the photon sphere.} use the lapse function $N$ as a coordinate in the spatial slice $\lbrace t=\mbox{const}\rbrace$. As Israel showed in \cite{Israel:1967}, the static vacuum equations (\ref{SME1}), (\ref{SME2}) can be rephrased as inequalities on each level set of the coordinate $N$. Integrating those inequalities from the photon sphere all the way to spatial infinity, we obtain inequalities relating the mean and Gau{\ss} curvatures of the photon sphere to the ADM-mass $m$ of the spacetime. Recall that we have already derived above that both of these curvatures are constant.

When combining the static vacuum equation (\ref{SME2}) with the asymptotic decay of the spatial metric $g$ discussed in (\ref{decay}) we can relate\footnote{This relationship can actually be pursued much further; it even constitutes a central tool for showing that the Newtonian limit of the ADM-mass of a static isolated spacetime ``is'' the Newtonian mass of ``its'' Newtonian limit. For more details, see \cite{Cederbaum:2011}.} the ADM-mass $m$ of the spacetime to the normal derivative $\nu(N)$ of the lapse function via
\begin{equation}\label{mass}
\frac{c^{2}}{4\pi G}\int_{\Sigma^{2}}\nu(N)\,d\sigma=m.
\end{equation}
This physical insight proves that the integrated inequalities are in fact equalities. This, however, implies that Israel's inequalities must be identities on each level set of $N$. From this, it is straightforward to compute that the spacetime is in fact Schwarzschild with $N$ playing the role of a rescaled radial variable. For more details, we refer the reader to \cite{Cederbaum:2014}.
 
\bibliographystyle{unsrt}

\bibliography{Cederbaum_eom_proceedings_2013}

\end{document}